\documentclass[journal]{IEEEtran}

\usepackage[T1]{fontenc}
\usepackage{cite}
\usepackage{amsmath,amssymb}
\usepackage{graphicx}
\usepackage{booktabs}
\usepackage{url}
\usepackage[hidelinks]{hyperref}

\begin{document}

\title{Routing Collapse in Multimodal Mixture-of-Experts:
Cross-Modal Conditioning Restores Expert Diversity
Without Improving Accuracy}

\author{Syed~Hussamuddin and
        Mubashra~Fayyaz%
\thanks{The authors are with the Department of Computer Science, FAST
National University of Computer and Emerging Sciences, Karachi 75030,
Pakistan (e-mail: syed.hussam.uddin1@gmail.com;
mubashra.fayyaz@nu.edu.pk).}%
\thanks{This work has been submitted to the IEEE for possible
publication. Copyright may be transferred without notice, after which
this version may no longer be accessible.}}

\markboth{IEEE Signal Processing Letters}%
{Hussamuddin \MakeLowercase{\textit{et al.}}: Routing Collapse in
Multimodal Mixture-of-Experts}

\maketitle

\begin{abstract}
Mixture-of-Experts (MoE) classifiers are increasingly applied to
multimodal emotion recognition, yet the health of their routing
distributions is rarely measured. We study a seven-expert MoE head on
IEMOCAP under two conditions that differ only in how modality
embeddings are fused before reaching it: a learned scalar gate alone,
and the same gate preceded by pairwise cross-modal attention. Under
scalar fusion the router collapses, concentrating on 1.88 of seven
experts at a normalised entropy of 0.241. Cross-modal fusion raises
entropy to 0.537 and utilisation to 3.25 experts, a $2.23\times$
increase ($p=0.035$, Cohen's $d=2.55$), and induces
emotion-differentiated routing absent in the scalar case. This
recovered diversity confers no detectable benefit: across six seeds per
condition, weighted accuracy ($p=0.646$) and run-to-run variance
($p=0.768$) are statistically indistinguishable. We further observe a
13-point accuracy spread under identical configuration. In this regime,
router health and task performance are decoupled, and diagnosing
collapse does not by itself justify fixing it.
\end{abstract}

\begin{IEEEkeywords}
Emotion recognition, mixture of experts, multimodal fusion,
reproducibility, routing collapse.
\end{IEEEkeywords}

\section{Introduction}

\IEEEPARstart{M}{ixture}-of-Experts (MoE) layers route each input to a
learned combination of parallel sub-networks. They have recently been
adopted for multimodal emotion recognition (MER) on the assumption that
different experts will specialise in different affective patterns
\cite{mengara2025,dutta2026,li2025summer}. The assumption is seldom
tested. Reported gains are attributed to expert specialisation without
measuring whether the router distributes load across experts at all.

This matters because MoE routers are known to collapse: the gate
converges onto a small subset of experts, reducing effective capacity
far below the nominal expert count while leaving parameter counts and
reported architecture unchanged. Collapse is well documented in
large-scale language modelling, where auxiliary load-balancing losses
are standard practice. It is rarely diagnosed in the small-data MER
regime, where fewer than four thousand training utterances must support
both routing and expert learning simultaneously.

We report a controlled study of this question. Holding the expert head
fixed, we vary only whether the fused representation it receives has
first been enriched by cross-modal attention, and we measure routing
behaviour directly rather than inferring it from accuracy. Our
contributions are:

\begin{enumerate}
\item We show that a seven-expert MoE head collapses to 1.88 effective
experts on IEMOCAP without cross-modal conditioning, with individual
runs reaching 1.53.
\item We show that cross-modal conditioning of the head's input
repairs this ($2.23\times$ entropy, $p=0.035$) and induces
emotion-differentiated expert assignment.
\item We show that the repair yields no detectable accuracy or
stability benefit over six seeds per condition.
\item We report a 13-point accuracy spread over six seeds under
identical configuration, indicating that single-run comparisons on this
benchmark are unreliable.
\end{enumerate}

\begin{figure*}[!t]
\centering
\includegraphics[width=\textwidth]{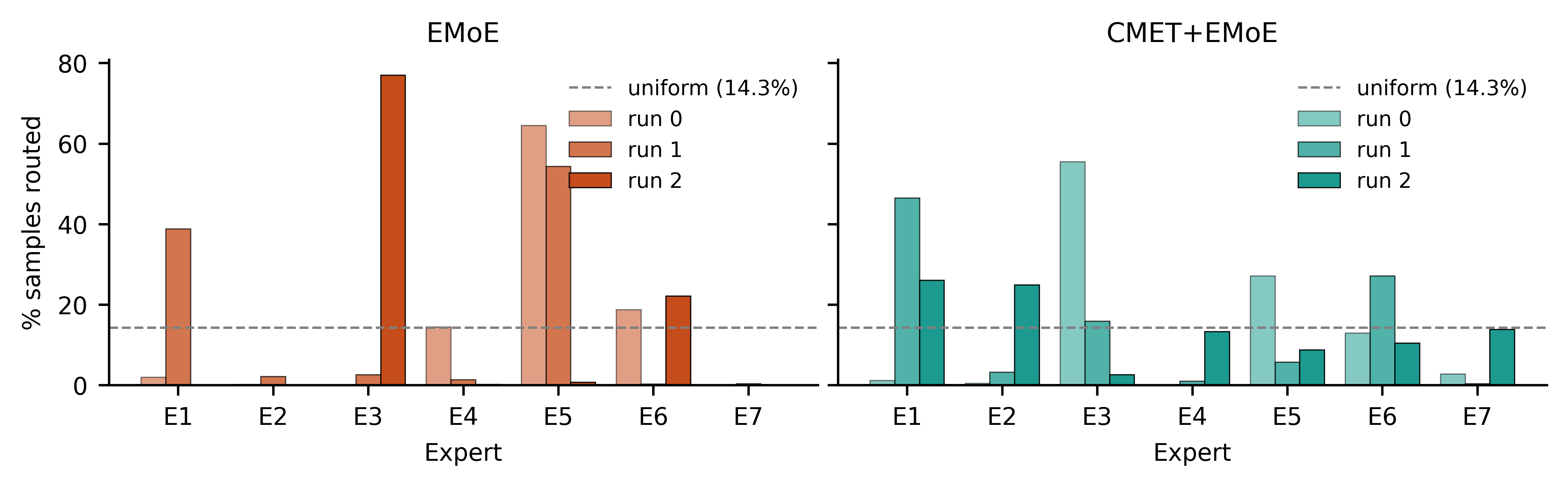}
\caption{Proportion of held-out samples for which each expert is the
argmax, over three seeds per condition (runs 0--2 are seeds 42--44).
Under EMoE (left) routing
concentrates on one or two experts, with several receiving no samples at
all; under CMET+EMoE (right) load is distributed more evenly. The dashed
line marks uniform routing (14.3\%).}
\label{fig:routing}
\end{figure*}

\section{Method}

\subsection{Backbone}

We use a trimodal encoder over the text, speech, and motion-capture
(MoCap) streams of IEMOCAP. Text is embedded with frozen
300-dimensional GloVe vectors \cite{pennington2014} and encoded by a
bidirectional LSTM followed by a Transformer block. Speech is
represented by 34 short-term features comprising MFCC, chroma, and
spectral descriptors, encoded by two Conv1D layers, a BiLSTM, and a
Transformer block. MoCap concatenates head, hand, and facial marker
streams into a $200\times189$ map encoded by three strided Conv2D
layers. Each branch emits a 128-dimensional embedding
$\mathbf{t},\mathbf{s},\mathbf{m}\in\mathbb{R}^{128}$. No pretrained
foundation model is used at any stage.

\subsection{Expert Head}

Both conditions share an identical seven-expert head. Given a fused
representation $\mathbf{f}\in\mathbb{R}^{256}$, a router assigns weights
and the prediction is their weighted combination:
\begin{equation}
\begin{aligned}
\mathbf{w}&=\mathrm{softmax}(\mathbf{W}_r\mathbf{f}+\mathbf{b}_r),
\qquad \mathbf{w}\in\mathbb{R}^{7},\\
\hat{\mathbf{y}}&=\mathrm{softmax}\Big(\textstyle\sum_{k=1}^{7}
w_k\,E_k(\mathbf{f})\Big),
\end{aligned}
\label{eq:moe}
\end{equation}
where each expert $E_k$ is a two-layer network mapping to four class
logits. No auxiliary load-balancing loss is applied. This is
deliberate: we wish to observe the routing behaviour the architecture
produces on its own, since MER work adopting MoE heads typically omits
such losses.

\subsection{The Two Conditions}

\textbf{EMoE.} The three branch embeddings are combined by a learned
scalar gate,
$\mathbf{f}_{\mathrm{gate}}=g_t\mathbf{t}+g_s\mathbf{s}+g_m\mathbf{m}$
with $\sum g=1$, and projected to $\mathbb{R}^{256}$.

\textbf{CMET+EMoE.} Before fusion, each modality attends to both others
through all six directed pairs using multi-head attention (4 heads,
$d_k=32$), a design related to prior gated cross-modal fusion for
emotion recognition \cite{lee2025,he2025} but computing the complete
pairwise graph rather than a single interaction. Writing
$\mathrm{cross}_x$ for the sum of the two cross-attention outputs
received by modality $x$, a learnable per-dimension gate combines own
and cross-modal content:
\begin{equation}
\begin{aligned}
\boldsymbol{\alpha}_x&=\mathrm{GELU}\big(\mathbf{W}_x[\mathbf{x};
\mathrm{cross}_x]+\mathbf{b}_x\big),\\
\mathbf{x}_{\mathrm{out}}&=\boldsymbol{\alpha}_x\odot\mathbf{x}
+(1-\boldsymbol{\alpha}_x)\odot\mathrm{cross}_x ,
\end{aligned}
\label{eq:gate}
\end{equation}
where $\odot$ denotes element-wise multiplication. Note that
$\boldsymbol{\alpha}_x$ is GELU-activated rather than sigmoid-activated
and is therefore not constrained to $[0,1]$: \eqref{eq:gate} is a
learned affine combination rather than a convex blend. The three outputs
are layer-normalised, concatenated, and passed through the same
$\mathbb{R}^{256}$ projection.

The conditions are otherwise identical and differ only in the fused
representation $\mathbf{f}$. Because $\mathbf{f}$ feeds both the router
and the experts in \eqref{eq:moe}, the comparison isolates the effect
of cross-modal fusion on the MoE head as a whole, not of the router
input alone. CMET also adds the parameters of its attention and gating
layers.

\begin{figure*}[!t]
\centering
\includegraphics[width=\textwidth]{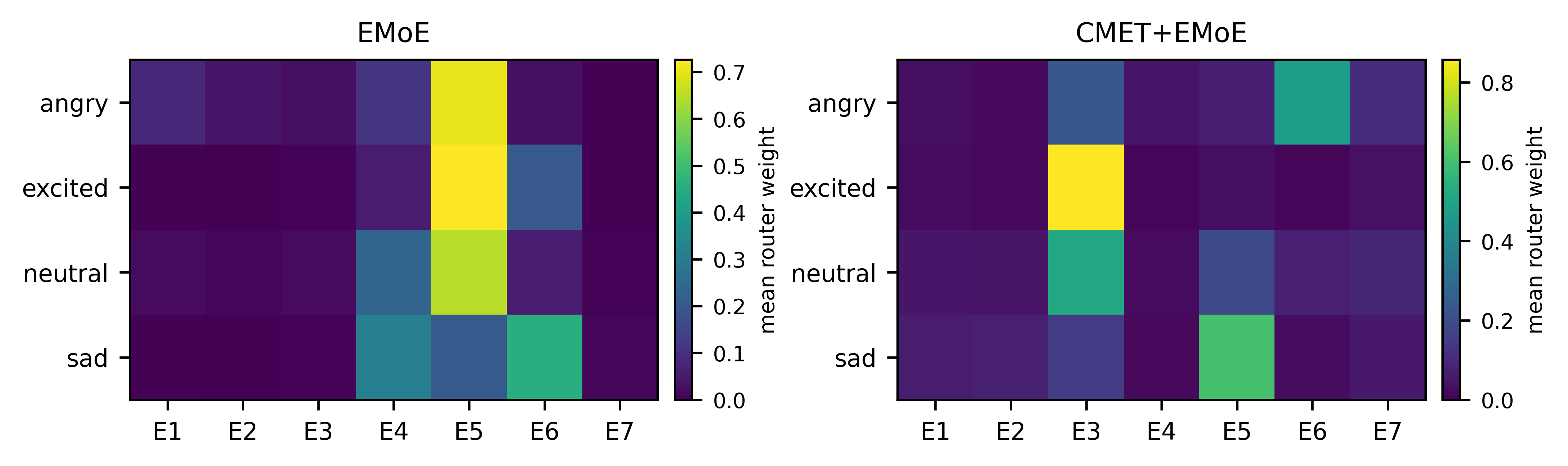}
\caption{Mean router weight per emotion class on the held-out session
for one seed per condition (seed 42, run 0 in Fig.~\ref{fig:routing}).
Expert indices are arbitrary and are not matched across seeds. Under
EMoE (left) three
of four emotions are dominated by a single expert. Under CMET+EMoE
(right) they are distributed across three distinct experts, indicating
emotion-differentiated routing.}
\label{fig:heatmap}
\end{figure*}

\section{Experimental Setup}

\subsection{Data}

We use a four-class subset of IEMOCAP \cite{busso2008}: anger
(1{,}103), sadness (1{,}084), neutral (1{,}708), and excitement
(1{,}041), for 4{,}936 utterances. The smaller happiness category,
which some prior work merges with excitement, is excluded. Sessions
1--4 form the training set (3{,}838 samples), and Session~5, whose
speakers are unseen in training, is held out for testing. We use this
single speaker-independent split rather than full five-fold
cross-validation so that six seeds per condition remain affordable,
since seed variance is itself a quantity of interest here.

\subsection{Training}

Both conditions use Adam with a cosine-decayed learning rate
($2\times10^{-3}\rightarrow10^{-6}$), categorical cross-entropy with
label smoothing 0.1, balanced class weights, batch size 64, and early
stopping with patience 12. Early stopping and checkpoint selection
monitored the held-out session, so absolute accuracies should be read
as optimistic; because both conditions share this procedure, it
favours neither.
Each condition is trained with \textbf{six random seeds} (42--47) under
otherwise identical configuration. Accuracy is reported over all six
runs. Routing statistics require the trained weights and are reported
over the \textbf{three runs} (seeds 42--44) for which checkpoints were
retained. Means are compared with two-sided Student's $t$-tests and
variances with a two-sided $F$-test.

\subsection{Routing Metrics}

For every run we record weighted accuracy (WA) and, on the held-out
session, the router output $\mathbf{w}_i$ for every test sample $i$. For
each sample we compute a normalised routing entropy
\begin{equation}
H_i=-\frac{1}{\log K}\sum_{k=1}^{K} w_{ik}\log w_{ik} ,
\qquad K=7,
\label{eq:entropy}
\end{equation}
where $H_i=0$ denotes total collapse onto a single expert and $H_i=1$
denotes uniform routing over all $K$. Both routing statistics are
reported as means over the $N$ test samples: the mean entropy
$\bar{H}=N^{-1}\sum_i H_i$, and the mean effective expert count
\begin{equation}
\bar{E}=\frac{1}{N}\sum_{i=1}^{N}\exp\!\big(H_i\log K\big),
\label{eq:eff}
\end{equation}
the average perplexity of the per-sample routing distribution. We note
that $\bar{E}\ge\exp(\bar{H}\log K)$ by Jensen's inequality, with the
gap growing in the spread of $H_i$; we average per-sample perplexities
rather than exponentiating the mean entropy, so that $\bar{E}$ reflects
how many experts are engaged for a typical input. We also report the
fraction of test samples for which each expert is the argmax, and the
number of experts receiving more than 1\% of samples.

\begin{table}[!t]
\centering
\caption{Results by condition (mean $\pm$ sd). Routing metrics are over
three seeds, accuracy over six. Routing differs significantly between
conditions; accuracy and its variance do not.}
\label{tab:main}
\setlength{\tabcolsep}{4pt}
\renewcommand{\arraystretch}{1.15}
\begin{tabular}{lccc}
\toprule
& \textbf{EMoE} & \textbf{CMET+EMoE} & \textbf{$p$} \\
\midrule
Router entropy $\bar{H}$ \textsuperscript{a}  & 0.241 $\pm$ 0.076 & \textbf{0.537 $\pm$ 0.145} & \textbf{0.035} \\
Effective experts $\bar{E}$ \textsuperscript{a}   & 1.88 $\pm$ 0.35   & \textbf{3.25 $\pm$ 0.76}   & \textbf{0.046} \\
Weighted accuracy \textsuperscript{b}   & 66.51 $\pm$ 4.98  & 67.79 $\pm$ 4.33           & 0.646 \\
Accuracy range \textsuperscript{b}      & 58.20--71.58      & 59.29--70.86      & --- \\
\bottomrule
\multicolumn{4}{l}{\footnotesize \textsuperscript{a}three seeds \quad
\textsuperscript{b}six seeds}
\end{tabular}
\end{table}

\section{Results}

\subsection{The Unconditioned Router Collapses}

EMoE attains a normalised entropy of $0.241\pm0.076$ and uses
$1.88\pm0.35$ of seven experts (Table~\ref{tab:main},
Fig.~\ref{fig:routing}). The behaviour is severe in individual runs.
In the most extreme, entropy falls to $0.155$ with $1.53$ effective
experts; two experts capture 99.1\% of test samples and three experts
receive no samples at all. Five of seven experts are, in effect, inert
parameters that contribute nothing to the decision while appearing in
the reported architecture.

\subsection{Cross-Modal Conditioning Repairs Routing}

With an identical expert head, CMET+EMoE reaches entropy
$0.537\pm0.145$ and $3.25\pm0.76$ effective experts. The entropy
increase of $2.23\times$ is significant at $p=0.035$ with a large effect
size (Cohen's $d=2.55$); the difference in effective experts is
significant at $p=0.046$.

The qualitative counterpart appears in the per-class routing profiles,
shown for a representative seed in Fig.~\ref{fig:heatmap}. Under EMoE, three of the four emotion
categories share a single dominant expert. Under CMET+EMoE the same four
categories are distributed across three distinct experts. Cross-modal
conditioning of the head's input is therefore sufficient to induce the
emotion-differentiated specialisation that such architectures are
designed to exhibit, and which does not otherwise emerge.

\subsection{The Repair Confers No Detectable Benefit}

Accuracy does not follow. WA is $66.51\pm4.98$ for EMoE and
$67.79\pm4.33$ for CMET+EMoE, a difference of $1.28$ points
($p=0.646$; 95\% CI $-4.7$ to $+7.3$). Run-to-run variance is likewise
indistinguishable ($F=1.32$, $p=0.768$), offering no support for the
intuition that a collapsed router destabilises training. At six seeds,
an accuracy gain larger than about seven points is excluded at 95\%
confidence, though a gain of a few points is not. The routing
distribution changed substantially; the task metric did not change
detectably.

\subsection{Entropy and Accuracy Are Decoupled}

The two conditions therefore separate on routing ($p=0.035$) but not
on accuracy ($p=0.646$). The same pattern appears within EMoE, though
over only three runs: its most collapsed run, at entropy $0.155$ with
$1.53$ effective experts, achieved the \emph{highest} accuracy of the
condition at 71.58\%, while runs with more diffuse routing did not
outperform it.

We interpret this as suggesting that the experts are not learning
complementary functions. If they were, distributing load across more
of them should help. That it does not suggests that, in this data
regime, spreading the training signal across additional experts
fragments it, and that the router's apparent specialisation is not
functional specialisation: the routing distribution differentiates
inputs, but the experts do not differentiate outputs.

\subsection{Seed Sensitivity}

Under identical configuration, EMoE accuracy ranges from 58.20\% to
71.58\% across six seeds, a spread of 13 points (11.6 points for
CMET+EMoE), and one seed degraded both conditions simultaneously. A
single run drawn from this distribution could support a wide range of
conclusions about the underlying architecture. Several frequently cited
IEMOCAP results report one run without a variance estimate
\cite{tripathi2018,khan2025}. If run-to-run variation is of similar
size in those settings, differences of one or two points between such
figures cannot be separated from seed effects.

\section{Discussion and Conclusion}

In this low-resource setting, the MoE router collapses without
intervention. Cross-modal conditioning of its input measurably repairs
this, raising entropy by $2.23\times$ and inducing
emotion-differentiated expert assignment, yet produces no detectable
accuracy or stability benefit.

Two implications follow for how MoE contributions in this literature are
evidenced. First, routing statistics should be reported directly.
Attributing gains to expert specialisation without measuring the routing
distribution is not supported by our results, since a head nominally
containing seven experts may be using fewer than two. Second, router
health should be treated as a diagnostic rather than an objective.
Interventions that demonstrably improve routing need not improve the
task, and load-balancing objectives adopted from large-scale language
modelling may not transfer to regimes where the experts lack sufficient
data to differentiate.

These findings come from a single benchmark, a single
speaker-independent split that was also used for model selection, a
fixed expert count, and routing statistics over three seeds. Whether the decoupling persists at larger data scales,
under full cross-validation, or with explicit load-balancing losses
remains open. Per-run routing measurements are available from the
authors on request.

\section*{Acknowledgment}

The authors used Claude (Anthropic) to help draft and edit the text of
this Letter. The research itself, including design, experiments,
measurement, statistical analysis, and interpretation, is the authors'
own. The authors verified all reported values and take full
responsibility for the content.

\end{document}